\documentclass[prd,aps,twocolumn,a4paper,floatfix,showpacs,nofootinbib]{revtex4-1}

\usepackage[utf8]{inputenc}  
\usepackage[T1]{fontenc}
\usepackage{graphicx,psfrag}
\graphicspath{{figures/}}
\usepackage{mathrsfs}
\usepackage{amsmath,amsfonts,amssymb}
\usepackage{multirow,enumerate}
\usepackage{comment,hyperref}
\usepackage{color}
\usepackage{acronym}
\usepackage{xspace}
\usepackage[normalem]{ulem}
\usepackage{mathtools}
\usepackage{subfigure}
\usepackage{makecell}
\usepackage{appendix}
\usepackage{diagbox}
\usepackage{lineno}

\newacro{BH}{black hole}
\newacro{NS}{neutron star}
\newacro{PN}{Post-Newtonian}
\newacro{BBH}{binary black hole}
\newacro{BNS}{binary neutron star}
\newacro{EOB}{effective-one-body}
\newacro{NR}{numerical relativity}
\newacro{GW}{gravitational wave}
\newacro{EOS}{equation-of-state}

\newcommand{\be}{\begin{equation}}
\newcommand{\ee}{\end{equation}}
\newcommand{\bea}{\begin{eqnarray}}
\newcommand{\eea}{\end{eqnarray}}
\newcommand{\bel}{\begin{align}}
\newcommand{\eel}{\end{align}}

\def\H{\mathcal{H}}
\def\m{\mu}
\def\n{\nu}

\def\GMc2{{\rm G M_{\odot} c^{-2}}}

\def\I{\mathcal{I}}

\def\SEOBNRv4T{\texttt{SEOBNRv4T}\xspace}

\usepackage{color}
\definecolor{cyan}{rgb}{0,0.9,0.9}
\definecolor{orange}{rgb}{0.9,0.5,0}
\definecolor{magenta}{rgb}{1,0,1}
\definecolor{purple}{rgb}{0.8,0.4,0.8}
\definecolor{gray}{rgb}{0.5,0.5,0.5}
\definecolor{mygreen}{rgb}{0.1,0.8,0.1}
\definecolor{darkblue}{rgb}{0.0,0.0,0.6}

\begin{document}

\title{Bounding dark charges on binary black holes using gravitational waves} 

\author{Pawan Kumar Gupta$^{1,2}$}
\author{Thomas F.M.~Spieksma$^{2}$}
\author{Peter~T.H.~Pang$^{1,2}$}
\author{Gideon Koekoek$^{1,3}$}
\author{Chris Van Den Broeck$^{1,2}$}

\affiliation{${}^1$Nikhef -- National Institute for Subatomic Physics, 
Science Park 105, 1098 XG Amsterdam, The Netherlands}
\affiliation{${}^2$Institute for Gravitational and Subatomic Physics (GRASP), 
Utrecht University, Princetonplein 1, 3584 CC Utrecht, The Netherlands}
\affiliation{${}^3$Department of Gravitational Waves and Fundamental Physics, 
Maastricht University, P.O.~Box 616, 6200 MD Maastricht, The Netherlands}

\date{\today}

\begin{abstract}
In models of minicharged dark matter associated with a hidden $U(1)$ symmetry, astrophysical 
black holes may acquire a ``dark'' charge, in such a way that the inspiral dynamics of binary black 
holes can be formally described by an Einstein-Maxwell theory. Charges enter 
the gravitational wave signal predominantly through a dipole term, but their effect
is known to effectively first post-Newtonian order in the phase, which enables  
measuring the size of the charge-to-mass ratios, $|q_i/m_i|$, $i = 1,2$, of the individual
black holes in a binary. We set up a Bayesian analysis to discover, or constrain, dark charges on binary 
black holes. After testing our framework in simulations, we apply it to selected 
binary black hole signals from the second Gravitational Wave Transient Catalog (GWTC-2), namely 
those with low masses so that most of the signal-to-noise ratio is in the inspiral
regime. We find no evidence for charges on the black holes, and place typical 1-$\sigma$ bounds on 
the charge-to-mass ratios of $|q_i/m_i| \lesssim 0.2 - 0.3$.  
\end{abstract}

\maketitle

\section{Introduction}
\label{sec:intro} 

The Advanced LIGO \cite{TheLIGOScientific:2014jea} and Advanced Virgo \cite{TheVirgo:2014hva} 
gravitational wave (GW) detectors have so far found more than 50 candidate signals \cite{gracedb},
the majority being from coalescing binary black holes 
\cite{Abbott:2016blz,Abbott:2016nmj,TheLIGOScientific:2016pea,LIGOScientific:2018mvr,Abbott:2020niy},
in addition to two binary neutron star inspirals \cite{TheLIGOScientific:2017qsa,Abbott:2020uma} 
and two neutron star-black hole events \cite{LIGOScientific:2021qlt}. 
In the analyses of the binary black hole signals, the sources were assumed to be  
well-modeled as pure vacuum spacetime, although tests of general relativity (GR) were performed
which allowed for deviations from the dynamics predicted by Einstein's theory 
\cite{TheLIGOScientific:2016src,LIGOScientific:2018jsj,Abbott:2018lct,LIGOScientific:2019fpa,LIGOScientific:2020tif}. 
In this work we will look into another possible source for modifications in binary 
black hole dynamics, namely electric charge.  

As is well-known, astrophysical black holes are unlikely to be able to accrue large amounts
of electric charge, at least in the context of the Standard Model of particle physics (see
e.g.~\cite{Cardoso:2016olt} for an overview). Kinematic build-up of charge through infall
of electrons is limited by the ratio of electron mass $m_e$ to charge $e$, to\footnote{In this paper
we use units such that $G = c = 4\pi\epsilon_0 = 1$, with $\epsilon_0$ the electric permittivity 
of the vacuum.} $Q/M \leq m_e/e \simeq  5 \times 10^{-22}$, with $Q$ and $M$ respectively 
the charge and mass of the black hole. Also dynamical processes such as charge accretion 
by a rotating black hole in a magnetic field $B$ can only produce charge-to-mass ratios 
of $Q/M \lesssim 1.7 \times 10^{-20}\,(M/M_\odot)\,(B/\mbox{Gauss})$ \cite{Wald:1974np}. 
Moreover, surrounding plasma in the form of interstellar matter will discharge even an extremal 
black hole with $Q = M$ on a timescale of $\tau \sim 10^{-6}$ s \cite{Eardley:1975kp}. 

The situation is different if one considers so-called minicharged dark matter models 
\cite{Holdom:1985ag,DeRujula:1989fe}, 
which involve new fermions that are charged under a hidden $U(1)$ symmetry and whose
``dark'' charges are much smaller than that of the electron. Such minicharged particles are  
viable cold dark matter candidates, and have been searched for in a variety of observations
and experiments 
\cite{Sigurdson:2004zp,Davidson:1993sj,Davidson:2000hf,McDermott:2010pa,Dubovsky:2003yn,Dolgov:2013una,Gies:2006ca,Gies:2006hv,Burrage:2009yz,Ahlers:2009kh,Haas:2014dda,Ball:2016zrp,Kadota:2016tqq}. 
For a dark fermion with mass $m$ and charge $q$, it is possible to have $m/q > 1$, in which case 
values of $Q/M \simeq 1$ can be attained, and discharge timescales by the 
surrounding (dark matter) plasma can be in the order of billions of years \cite{Cardoso:2016olt}. 

Assuming a single dark fermion and dark photon, the interaction of the hidden sector with gravity
can be described by an Einstein-Maxwell action 
\be
S = \int d^4x\,\sqrt{-g}\,\left[\frac{R}{16\pi} - \frac{1}{4} F_{\m\n} F^{\m\n} + 4\pi A_\m j^\m \right],
\label{eq:action}
\ee
with $g$ the determinant of the metric $g_{\m\n}$, $A_\m$ the vector potential of the hidden $U(1)$ interaction, 
$F_{\m\n} = \nabla_\m A_\n - \nabla_\n A_\m$ the associated field tensor, and 
$j^\m$ the hidden current. Here we want to look at the inspiral of binary black holes
in the presence of a dark sector, and search for, or put bounds on, dark charges which may be 
carried by them, using some of the GW signals that have been detected. The leading 
post-Newtonian modification to the phase is at -1PN in the usual notation, corresponding 
to dipole radiation. This is mostly determined by the combination
\be
\xi = \left| \frac{q_1}{m_1} - \frac{q_2}{m_2} \right|,
\label{eq:zeta}
\ee
where $(q_1, q_2)$ and $(m_1, m_2)$ are respectively the charges and masses of the individual 
black holes \cite{Barausse:2016eii,Cardoso:2016olt}. However, Khalil et al.~\cite{Khalil:2018aaj} 
also computed higher-order effects, at 0PN and 1PN orders in phase, in the context of 
Einstein-Maxwell-dilaton theory, which reduces to Einstein-Maxwell theory when scalar 
charges are set to zero. Since these beyond-leading order contributions also depend on different 
combinations of $q_1/m_1$ and $q_2/m_2$ from the one in Eq.~(\ref{eq:zeta}), including them will 
allow us to make statements on these two quantities separately. Thus, our gravitational waveform model will
include these modifications to the point particle inspiral phase, in addition to effects of 
(precessing) spins, which start from 1.5PN order. Finally, though leading-order 
modifications of the ringdown spectrum of the remnant black hole resulting from the merger
have been computed \cite{Pani:2013ija,Pani:2013wsa,Zilhao:2014wqa,Mark:2014aja,Dias:2015wqa},
here we will focus only on the post-Newtonian inspiral, since to our knowledge the behavior at 
plunge and merger, which connects the early inspiral to the ringdown, has 
yet to be analytically investigated in the presence of charge. We will be particularly 
interested in relatively low-mass binary black hole signals, for which inspiral dominates
the signal-to-noise ratio.

This paper is structured as follows. In Sec.~\ref{sec:framework} we explain our 
waveform approximant and the data analysis set-up. In Sec.~\ref{sec:simulations} we
describe simulations that were done to provide a basic validation of the analysis framework.
Sec.~\ref{sec:realsignals} applies our methodology to a selection of detected signals. 
A summary and conclusions are provided in Sec.~\ref{sec:conclusions}.

\section{Waveform model and analysis framework}
\label{sec:framework}

Our baseline waveform model will be the frequency domain inspiral-merger-ringdown 
approximant IMRPhenomPv2 \cite{hannam:2013oca,Husa:2015iqa,PhysRevD.93.044007}, which we
modify to reflect the presence of charges. This waveform 
stitches together in C$^1$ fashion 
(a) an \emph{inspiral} regime which mostly follows the post-Newtonian description; 
(b) a phenomenological \emph{intermediate} regime 
describing the late inspiral and plunge; and (c) a phenomenological \emph{merger-ringdown} regime. 
Spin precession is captured by ``twisting up'' an underlying aligned-spin model
\cite{Schmidt:2012rh,Schmidt:2014iyl}. Since with current detectors most of our information
tends to come from the phase rather than the amplitude (though also see 
\cite{LIGOScientific:2020stg,Abbott:2020khf}), we will focus on the former, and in particular
on the inspiral phase. When electric charges are small and the inspiral is mainly driven by 
the tensor quadrupole flux, a good approximation to the inspiral phase is given by
\begin{widetext}
\be
\phi_{\rm Ins}(v) = 2\pi f t_c - \varphi_c - \pi/4 
+ \frac{1}{v^5}\left[ \frac{\rho^{\rm QD}_{-2}}{v^2} + \rho^{\rm QD}_0 + \rho^{\rm QD}_2 v^2 
+ \phi_{\rm Ins}^{\rm higher-order}(v) \right]. 
\label{eq:phase}
\ee
\end{widetext}
Here $t_c$ and $\varphi_c$ are respectively a reference time and reference phase. 
One has $v = (\pi M f)^{1/3}$, with $f$ the GW frequency, and where $M$ is a ``dressed'' total mass; 
specifically 
$M = G_{12} \bar{M}$, with $G_{12} = 1 - q_1 q_2/(m_1 m_2)$, where $\bar{M}$ is the observed total mass 
in the absence of charges. The first three terms in the square brackets include the  
charge-induced modifications to the phase computed by Khalil et al.~\cite{Khalil:2018aaj} up to 
1PN order. The leading-order (-1PN) contribution is set by 
\be
\rho^{\rm QD}_{-2} = -\frac{5G_{12}}{3584\nu} 
\left( \frac{q_1}{m_1} - \frac{q_2}{m_2} \right) ^2,
\label{eq:-1PN}
\ee
with $\nu = m_1 m_2/M^2$ the symmetric mass ratio. 
The expressions for $\rho^{\rm QD}_0$ and $\rho^{\rm QD}_2$ will not be shown explicitly here, since
they are obtained straightforwardly from the ones in \cite{Khalil:2018aaj} (see 
their Eqs.~(3.34a)-(3.34c) and Appendix B) by setting scalar charges to zero 
but retaining electric charges. These coefficients further reduce to the usual 0PN and 1PN 
coefficients for the vacuum case when electric charges are also set to zero. 
Finally, $\phi_{\rm Ins}^{\rm higher-order}$ collects all higher-order contributions in $v$, including
PN contributions as well as phenomenological corrections to the late inspiral, as detailed in 
\cite{Husa:2015iqa}. 

In the IMRPhenomPv2 approximant, the inspiral regime is stitched onto the intermediate 
regime at a frequency $f$ such that $M f = 0.018$. Since we do not know how charges affect 
the latter regime, one option would be to smoothly let the waveform go to zero around that 
frequency, e.g.~by applying a Planck tapering window \cite{McKechan:2010kp}. However, 
especially when performing parameter estimation on high-mass systems for which the merger
is well inside the detectors' sensitive band, we found a tendency for the tapered template waveform
to try and match part of the post-inspiral signal, leading to a significant underestimation 
of the masses. As a pragmatic solution, we opt to not taper the waveform; instead we will 
only analyze signals for which less than 5\% of the matched-filtering signal-to-noise ratio 
is contained in the regime $M f > 0.018$. Note that this transition always precedes 
the nominal last stable orbit (given by $M f_{\rm LSO} = 1/(6^{3/2}\pi) \simeq 0.022$), 
so that in this way we select signals for which only the inspiral has significant power in band.

Next we turn to our data analysis framework. 
Given a detected binary black hole coalescence signal, a waveform approximant $\tilde{h}_{\rm C}(f)$ 
with modified phasing as in Eq.~(\ref{eq:phase}) can be viewed as corresponding to a Bayesian hypothesis 
$\H_{\rm C}$, which states that one or both of the black holes carried a Maxwell charge. If on
the other hand charges are restricted to zero, the associated waveform model $\tilde{h}_{\rm NC}$
defines a hypothesis $\H_{\rm NC}$, stating that no charges were present. Given a hypothesis $\H$,
data $d$, and whatever background information $\I$ we may possess, a Bayesian evidence is
obtained through \cite{Veitch:2009hd}
\be
p(d | \H, \I) = \int d\bar\theta \,p(d | \H, \bar\theta, \I)\,p(\bar\theta | \H, \I).
\label{eq:evidence}
\ee
The integral is over the parameters $\bar\theta$ (masses, spins, possibly electric charges, ...)
that enter the waveform model $\tilde{h}(\bar\theta; f)$ associated with $\H$. $p(\bar\theta|\H, \I)$
is the prior density, and the likelihood $p(d | \H, \bar\theta, \I)$ is given by 
\be
p(d | \H, \bar\theta, \I) 
\propto \exp\left[ -\langle d - h(\bar\theta) | d - h(\bar\theta) \rangle/2 \right].
\label{eq:likelihood}
\ee
The noise-weighted inner product $\langle \, \cdot \, | \, \cdot \, \rangle$ is defined as
\be
\langle a | b \rangle 
= 4\Re \int_{f_{\rm low}}^{f_{\rm high}} df\,\frac{\tilde{a}^\ast(f)\,\tilde{b}(f)}{S_n(f)},
\ee
where $f_{\rm low}$ and $f_{\rm high}$ are respectively the detector's lower cut-off frequency
and the frequency at which a given signal ends, and $S_n$ is the (one-sided) power spectral density of
the noise. In this paper we will set $f_{\rm low} = 20$ Hz, and $f_{\rm high}$ is determined
by the parameters entering the IMRPhenomPv2 waveform. 
Given our hypotheses $\H_{\rm C}$ and $\H_{\rm NC}$, the general expression 
for the evidence (\ref{eq:evidence}) enables computation of a Bayes factor which can be 
used to rank the hypotheses:
\be
B^{\rm C}_{\rm NC} \equiv \frac{p(d|\H_{\rm C}, \I)}{p(d|\H_{\rm NC}, \I)}.
\ee

Apart from hypothesis ranking, the Bayesian framework also allows us to perform 
parameter estimation; in particular, the posterior density function (PDF) for the 
parameters $\bar\theta$ follows from Bayes' theorem:
\be
p(\bar\theta | \H, d, \I) 
= \frac{p(d | \H, \bar\theta, \I)\,p(\bar\theta | \H, \I)}{p(d | \H, \I)}.
\ee
The one-dimensional PDF $p(\lambda | \H, d, \I)$ for a particular parameter $\lambda$
is obtained from this by integrating out all other parameters. 

In our studies, the likelihood function of Eq.~(\ref{eq:likelihood}) is sampled using
the \texttt{lalinference\_nest} algorithm in the LALInference library \cite{Veitch:2014wba}, 
while the waveform model described in the previous section was implemented as an extension 
of the IMRPhenomPv2 approximant in the LALSimulation library of LALSuite \cite{lalsuite}. 

Together with spin-related parameters, the intrinsic parameters being sampled over directly are the 
(dressed) total mass $M$, the mass ratio $q = m_2/m_1$ (with the convention $m_2 \leq m_1$), 
and the charge-to-mass ratios $\sigma_1 \equiv q_1/m_1$, 
$\sigma_2 \equiv q_2/m_2$. For $M$ and $q$ we use uniform priors chosen wide enough so as to
accommodate the supports of the PDFs (with an upper bound of 1 for $q$). Regarding the $\sigma_i$, 
for the examples in this paper a uniform prior spanning $\sigma_i \in [-2,2]$ amply sufficed; 
here the sampling was done with the additional constraint $\sigma_1 \sigma_2 < 1$, corresponding 
to the requirement of inspiraling orbits. Priors on the spin magnitudes $a_1$, $a_2$ are taken to be 
uniform in the range $[0, 0.99]$, and priors on spin directions are uniform on the sphere. Both for
simulated signals and for template waveforms we impose the Kerr-Newman condition for the 
presence of black hole horizons, i.e.~$a_i^2 + \sigma_i^2 \leq 1$, $i =1, 2$ \cite{Newman:1965my}.  
For the extrinsic parameters, the priors on sky position and the orientation of 
the orbital plane at some reference frequency are also uniform on the sphere. We use a uniform-in-volume prior on distance, up to a maximum distance needed to accommodate the PDF.

\section{Simulations}
\label{sec:simulations}

\begin{figure*}
    \centering
   \includegraphics[keepaspectratio, width=12.0cm]{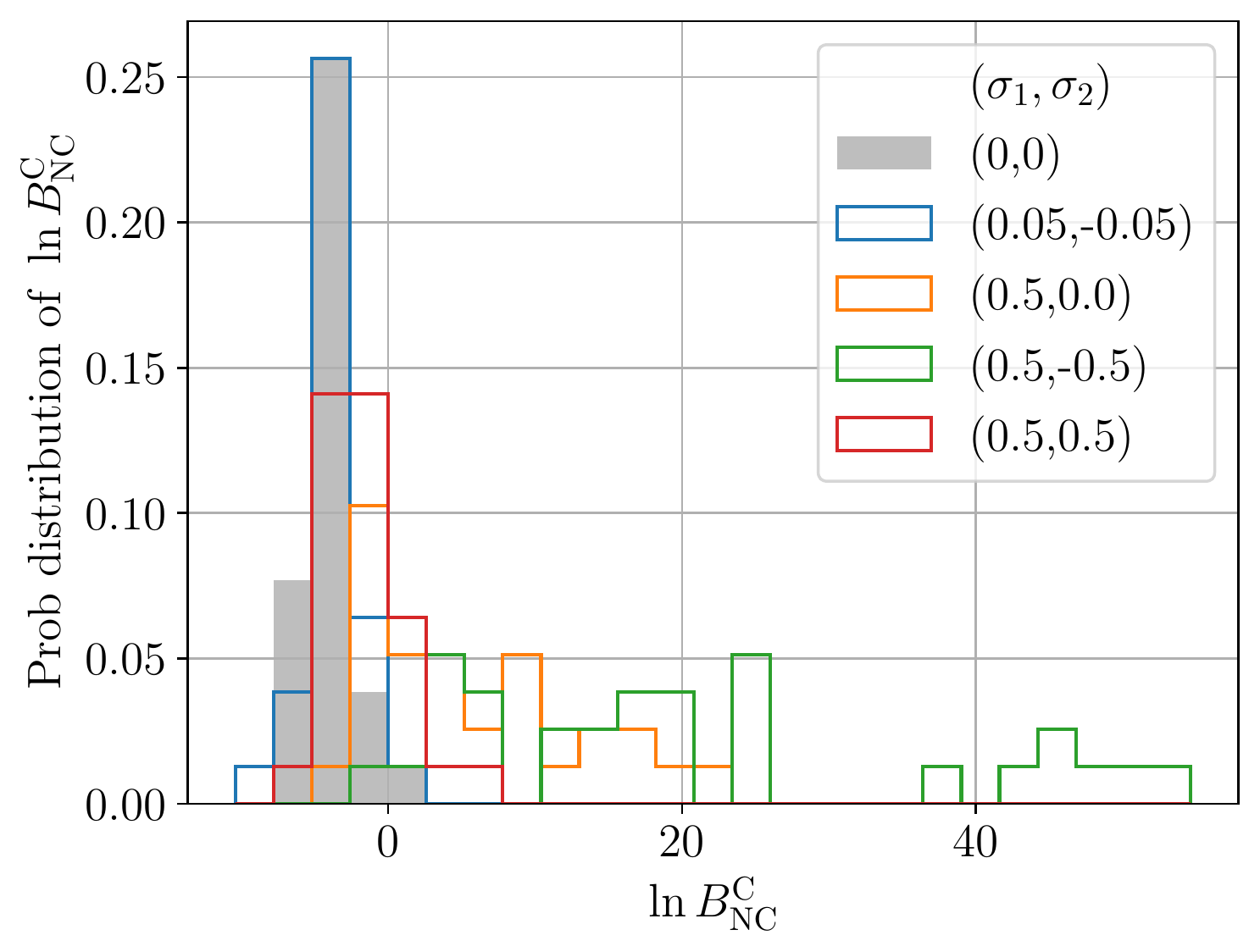}
    \caption[width = 0.8\textwidth]{Histograms for $\ln B^{\rm C}_{\rm NC}$ for 67 choices
    of masses and spins with ranges as detailed in the main text, and the five different choices of 
    $(\sigma_1, \sigma_2)$ indicated in the legend. 
    }  
    \label{fig:histograms}
\end{figure*}

\begin{figure*}
    \centering
   \includegraphics[keepaspectratio, width=8.5cm]{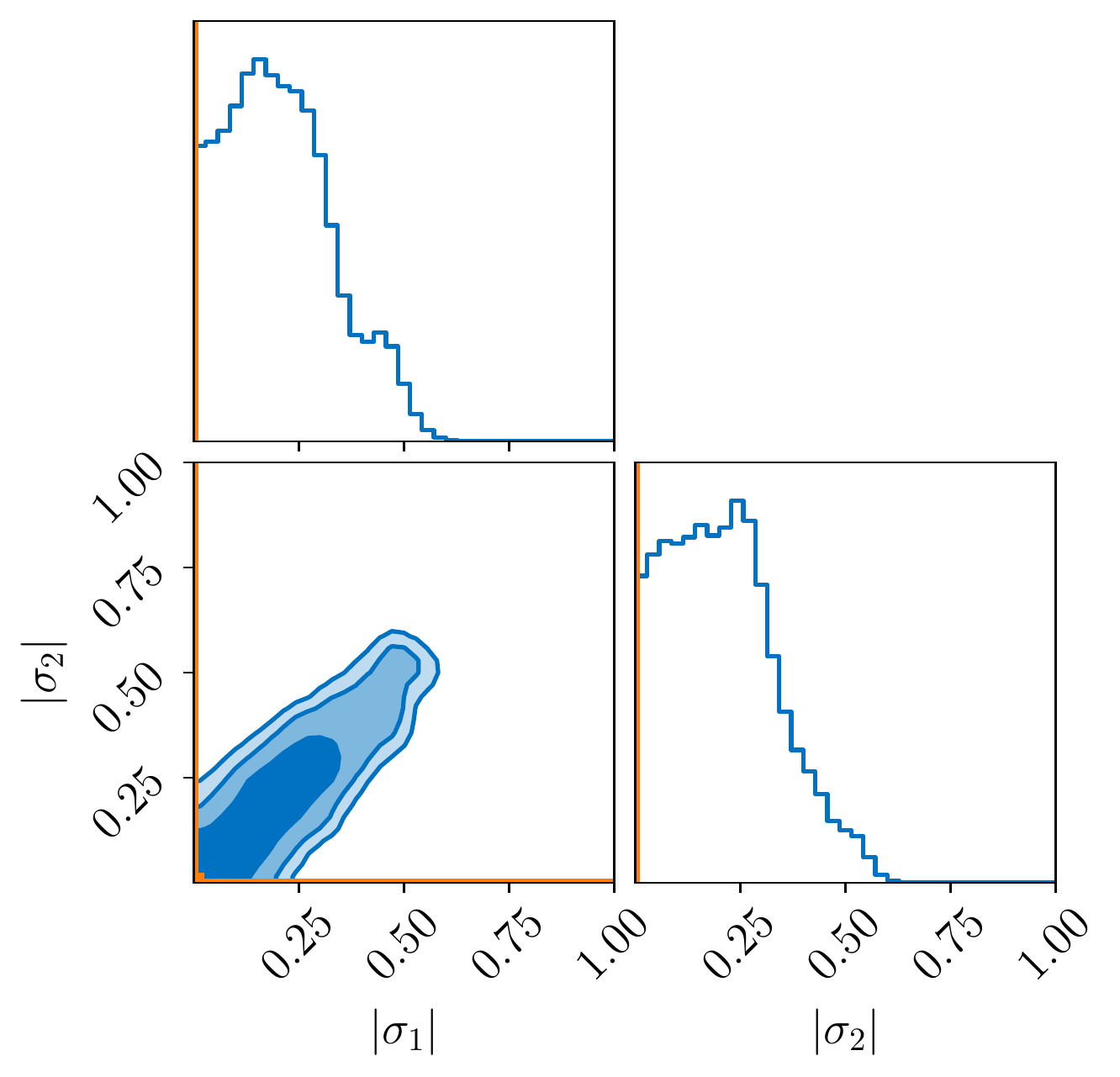}
   \includegraphics[keepaspectratio, width=8.5cm]{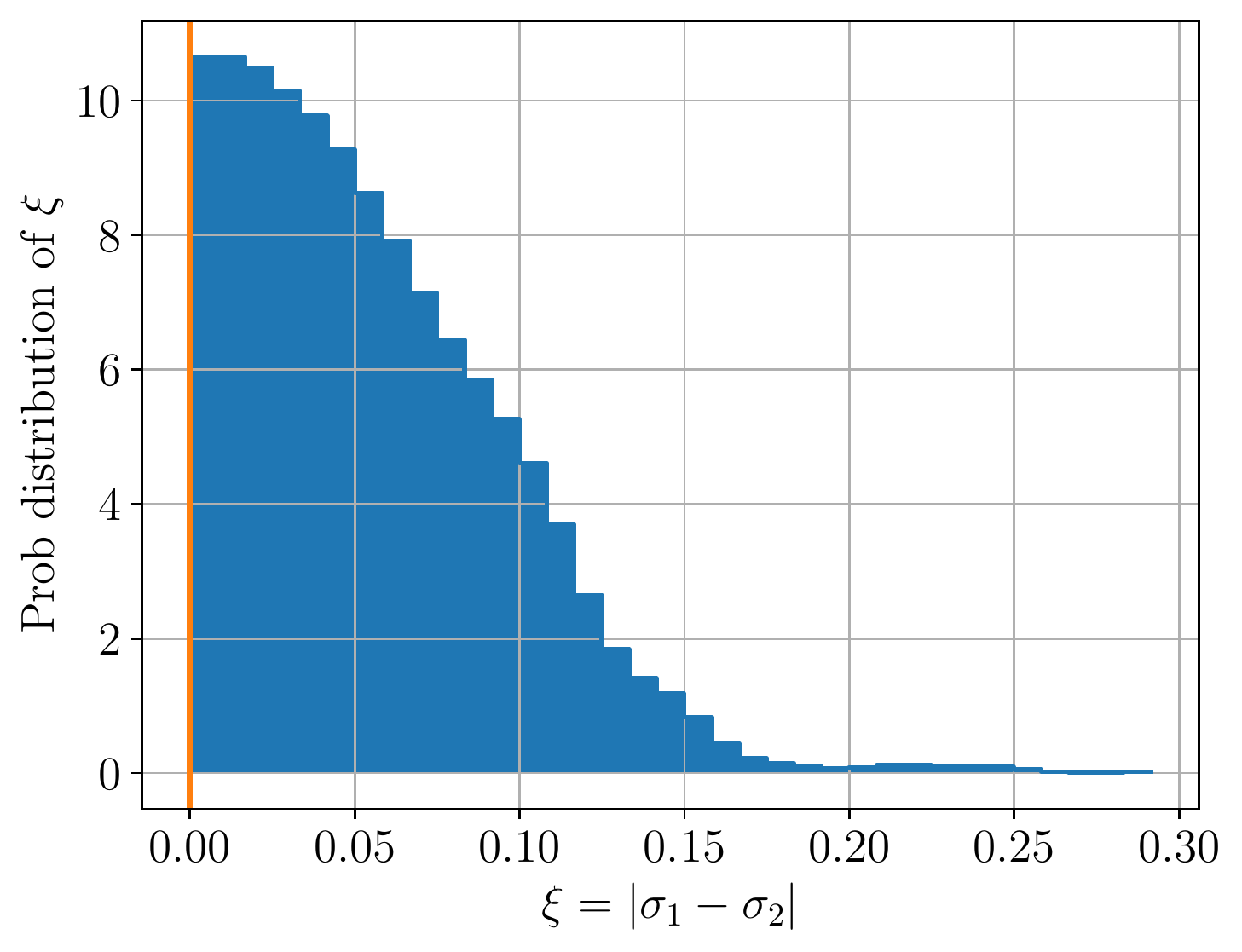}
    \caption[width = 0.8\textwidth]{Posterior distributions for an injection with 
    $(m_1, m_2) = (13.87, 6.36)\,M_\odot$ at an SNR of 12.52, and $(\sigma_1, \sigma_2) = (0,0)$. The left panel shows a 
    corner plot for the posterior distributions of $|\sigma_1|$ and $|\sigma_2|$ (with the contours
    enclosing respectively 68\%, 95\%, and 99.7\% of probability), while the right one
    is the posterior for $\xi = |\sigma_1 - \sigma_2|$. Here and in the analogous figures below, 
    orange lines indicate the injected parameters.
    }  
    \label{fig:00}
\end{figure*}

\begin{figure*}
    \centering
   \includegraphics[keepaspectratio, width=8.5cm]{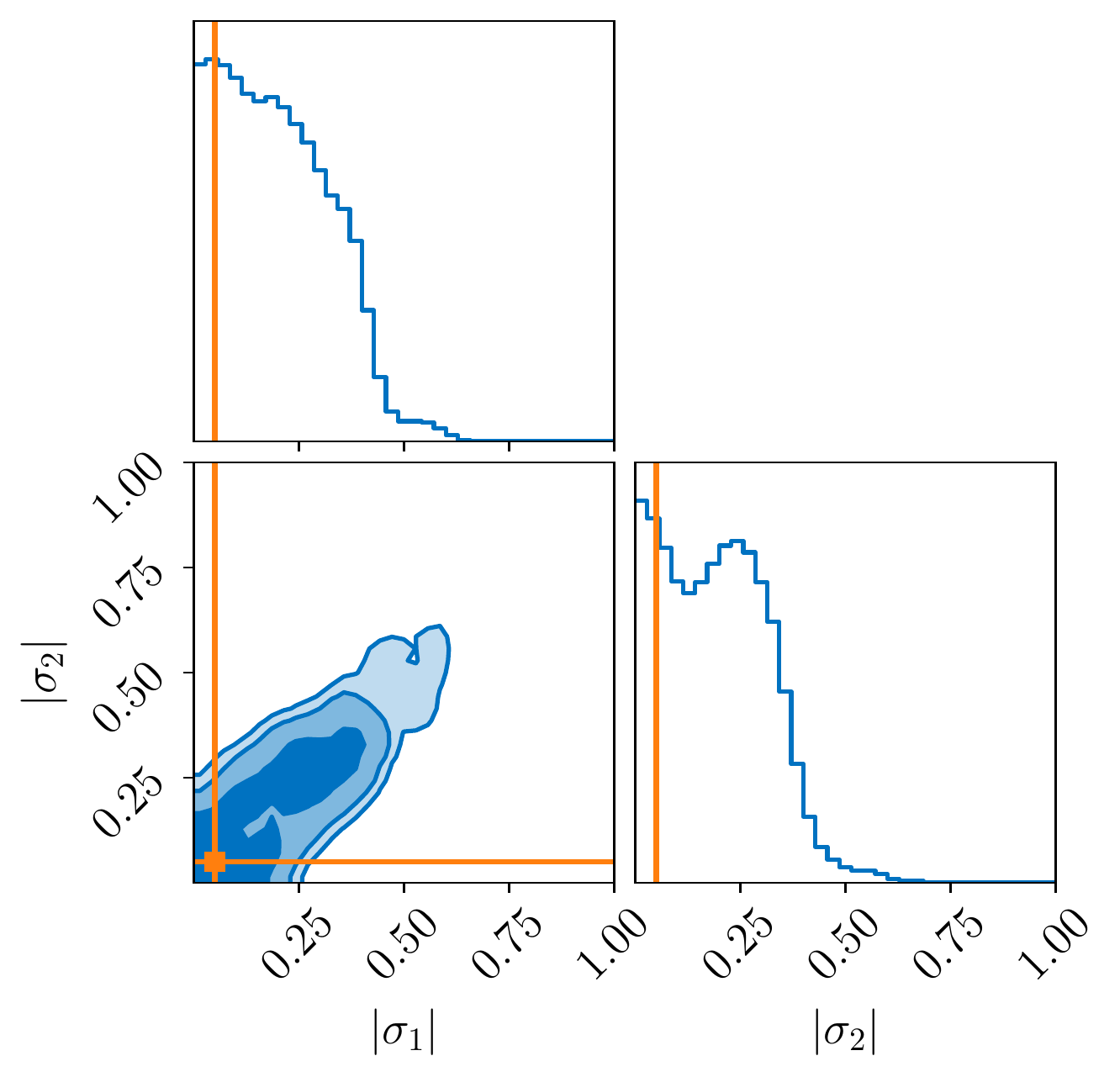}
   \includegraphics[keepaspectratio, width=8.5cm]{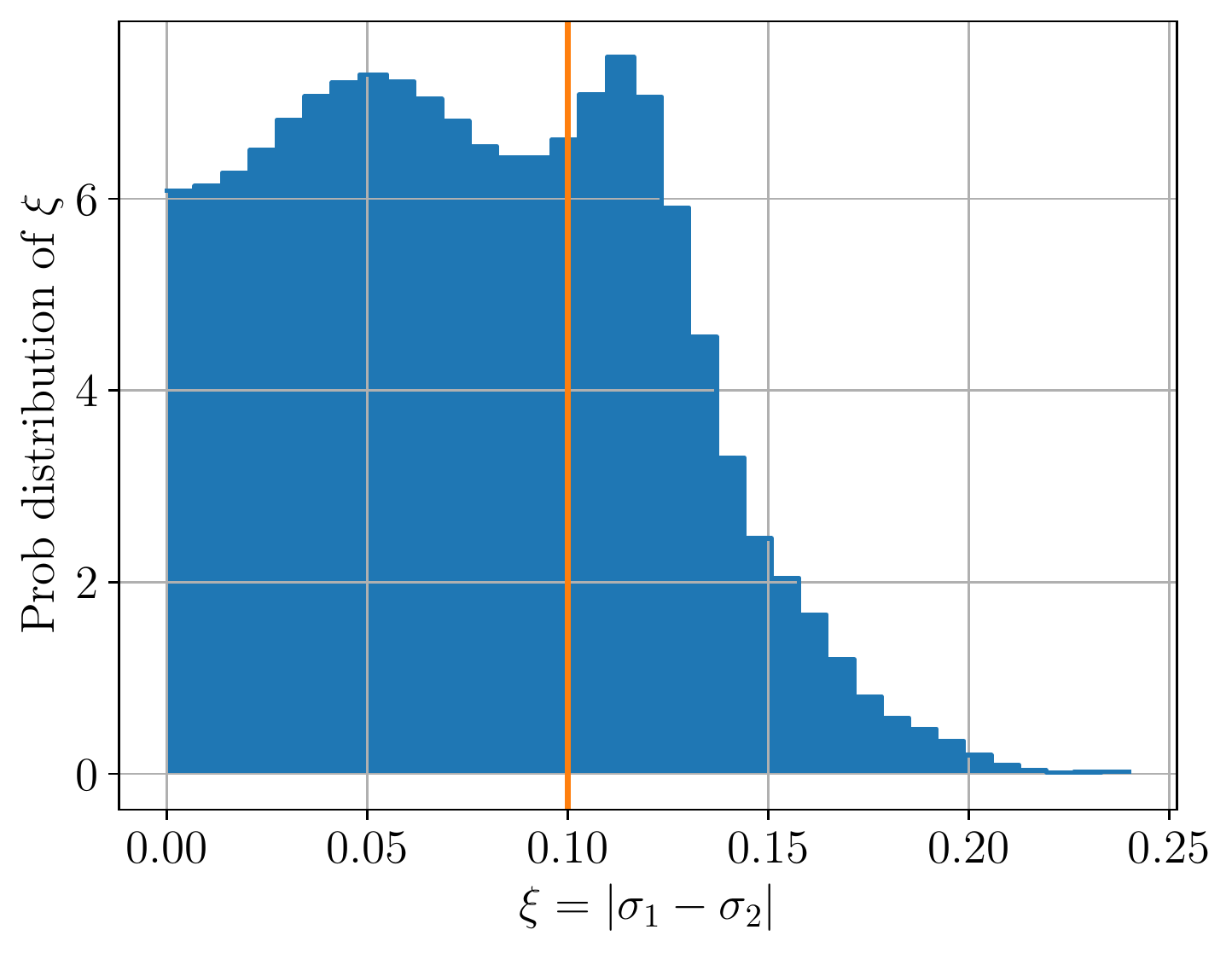}
    \caption[width = 0.8\textwidth]{Same as in Fig.~\ref{fig:00} but for 
    $(\sigma_1, \sigma_2) = (0.05, -0.05)$. 
    }  
    \label{fig:0p05n0p05}
\end{figure*}

\begin{figure*}
    \centering
   \includegraphics[keepaspectratio, width=8.5cm]{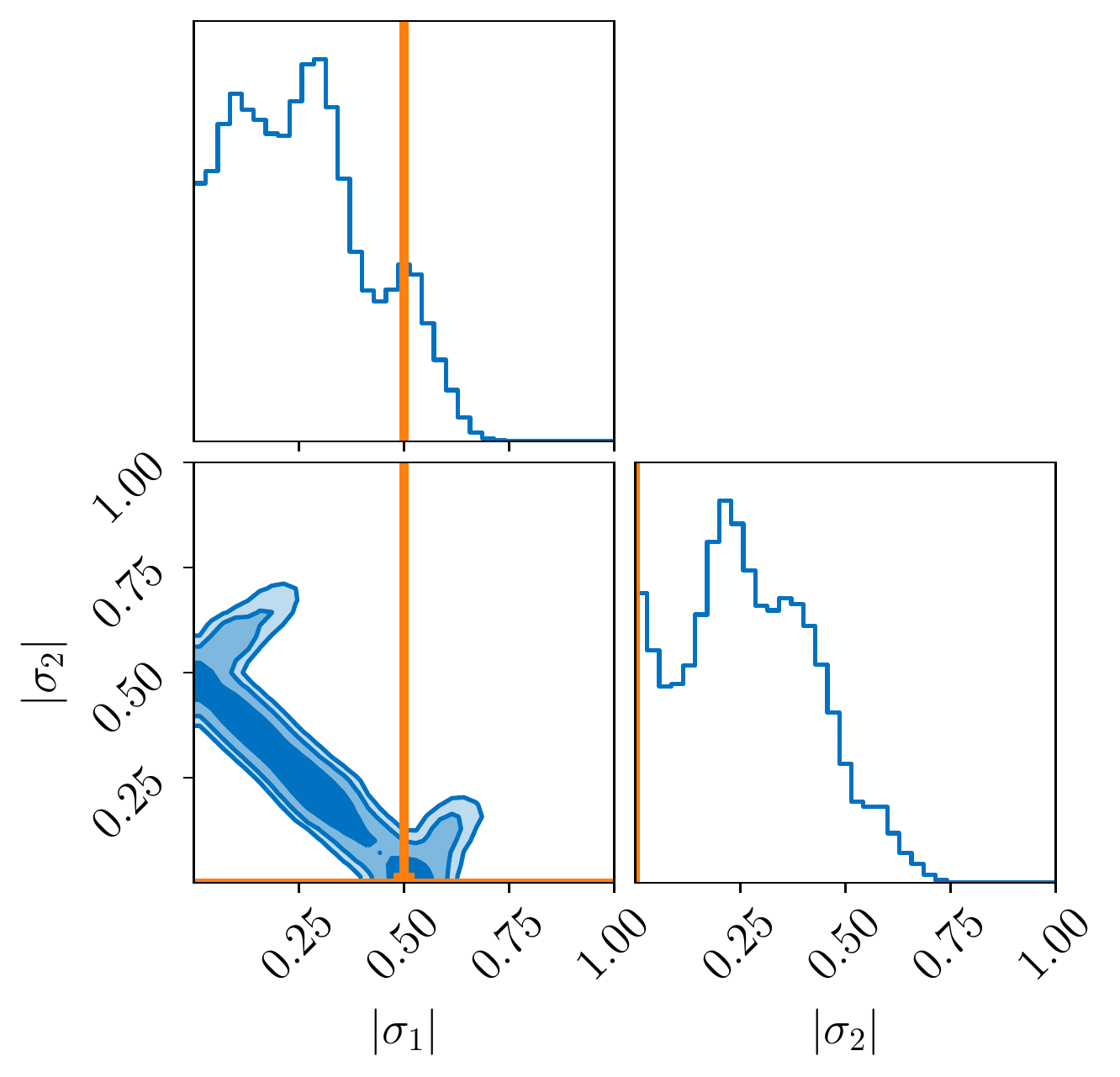}
   \includegraphics[keepaspectratio, width=8.5cm]{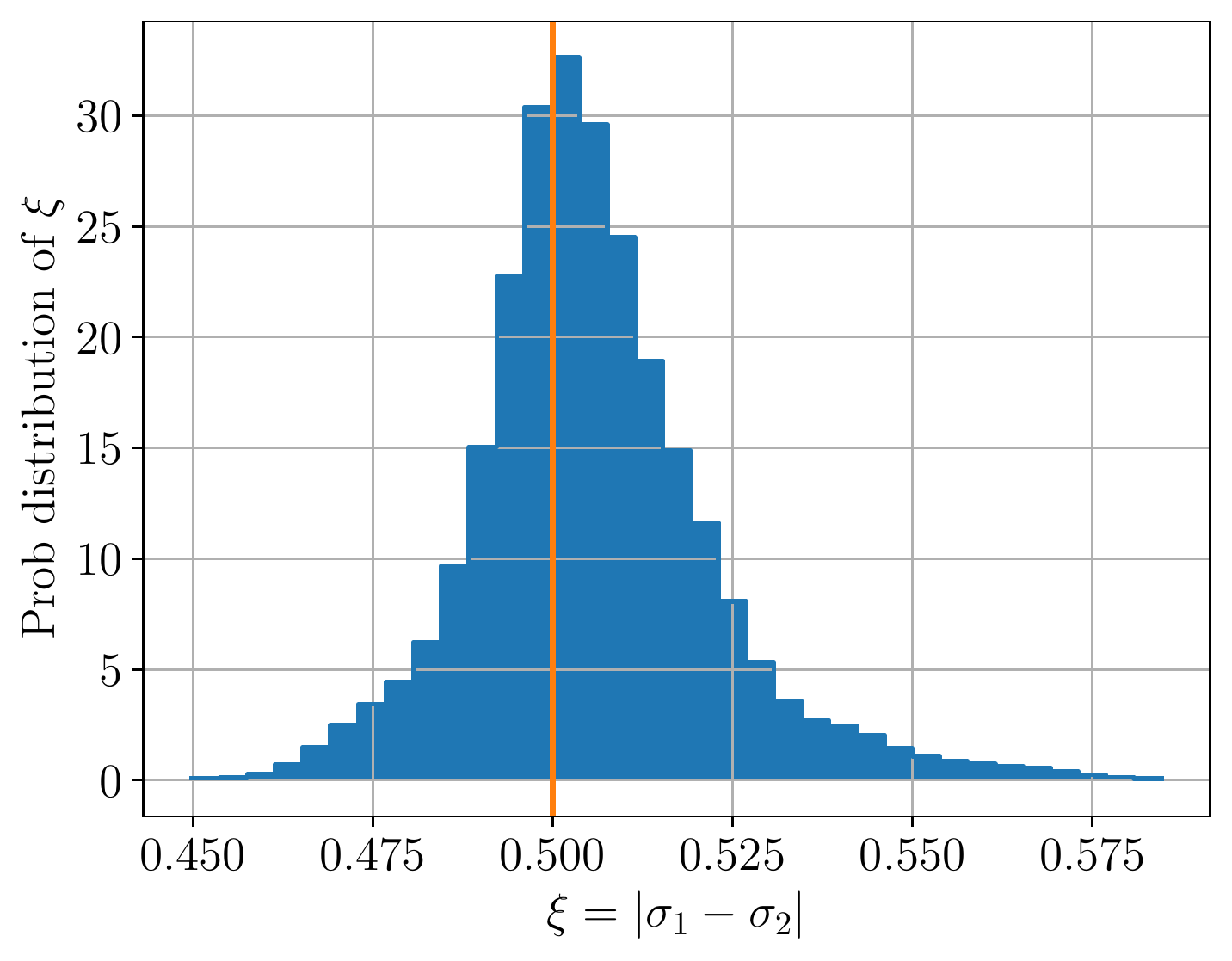}
    \caption[width = 0.8\textwidth]{Same as in Fig.~\ref{fig:00} but for 
    $(\sigma_1, \sigma_2) = (0.5, 0)$. 
    }  
    \label{fig:0p50}
\end{figure*}

\begin{figure*}
    \centering
   \includegraphics[keepaspectratio, width=8.5cm]{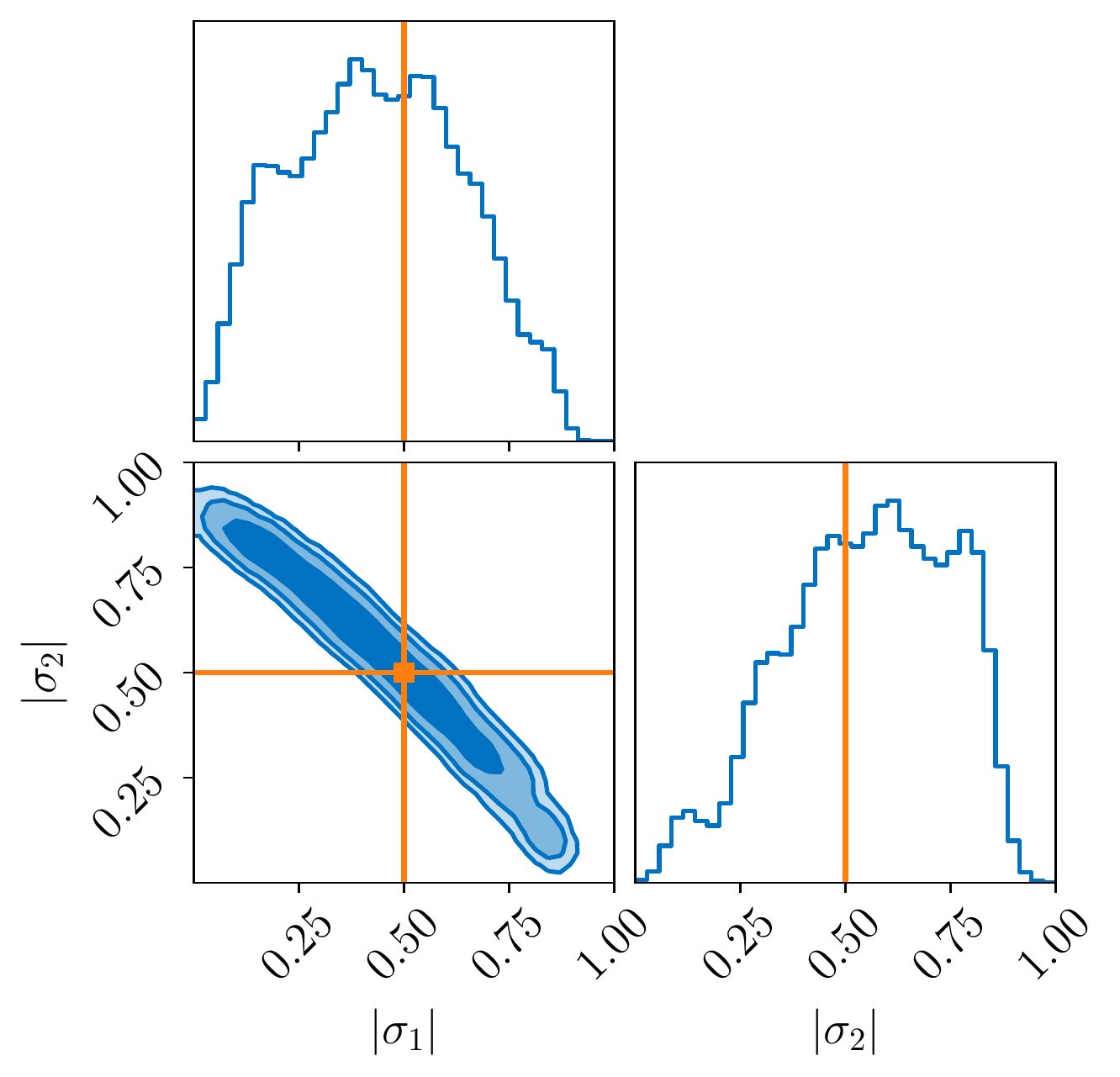}
   \includegraphics[keepaspectratio, width=8.5cm]{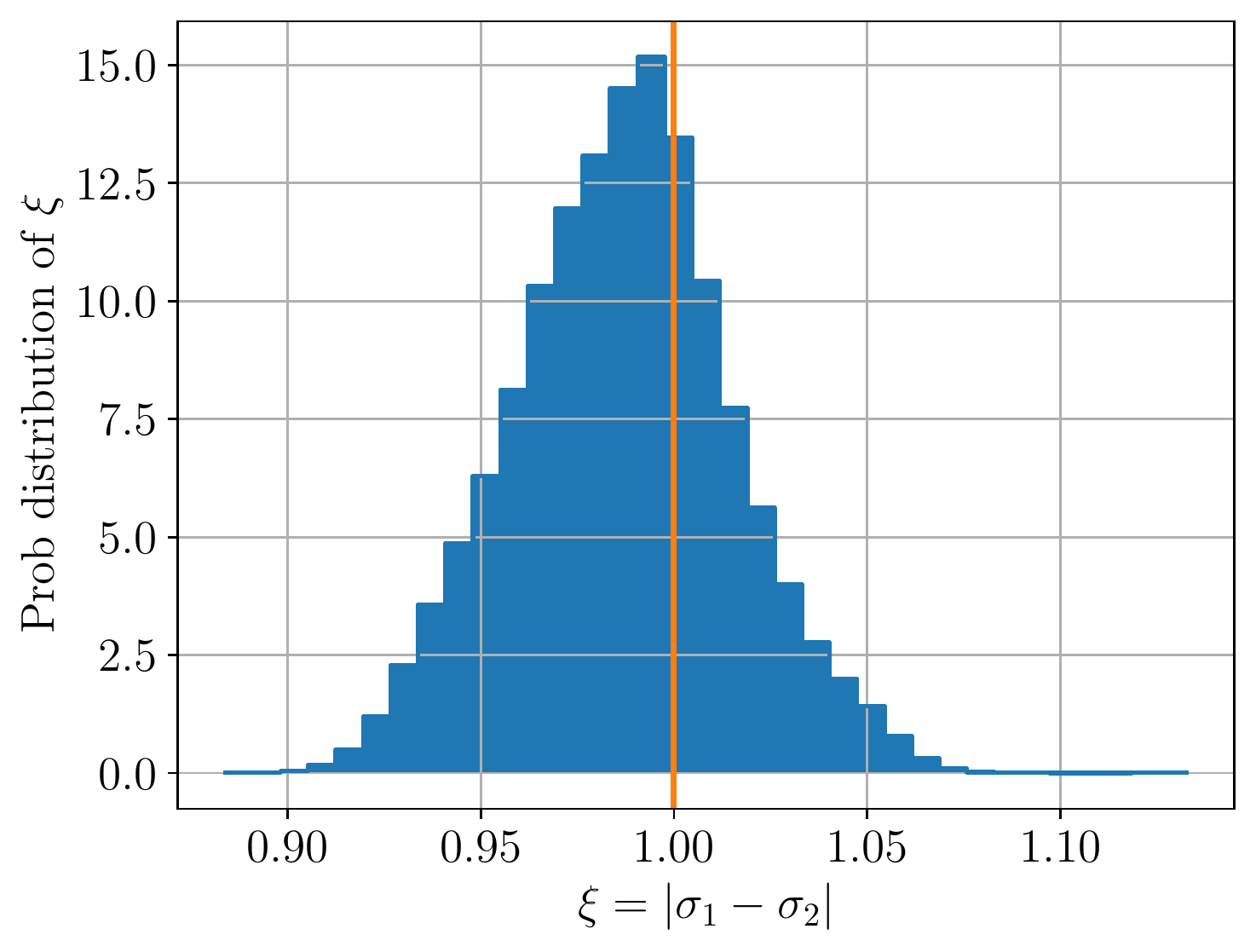}
    \caption[width = 0.8\textwidth]{Same as in Fig.~\ref{fig:00} but for 
    $(\sigma_1, \sigma_2) = (0.5, -0.5)$. 
    }  
    \label{fig:0p5n0p5}
\end{figure*}

\begin{figure*}
    \centering
   \includegraphics[keepaspectratio, width=8.5cm]{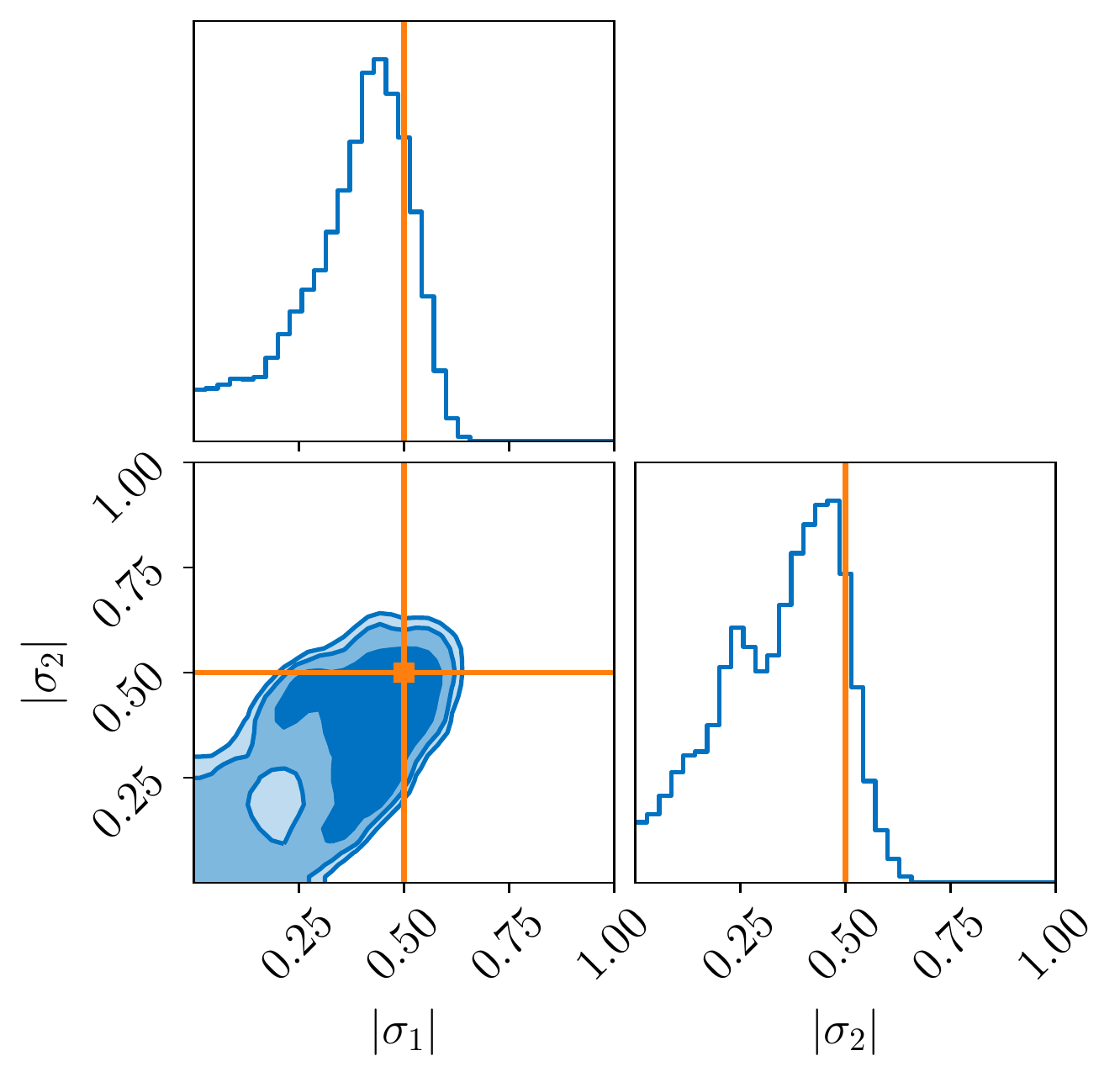}
   \includegraphics[keepaspectratio, width=8.5cm]{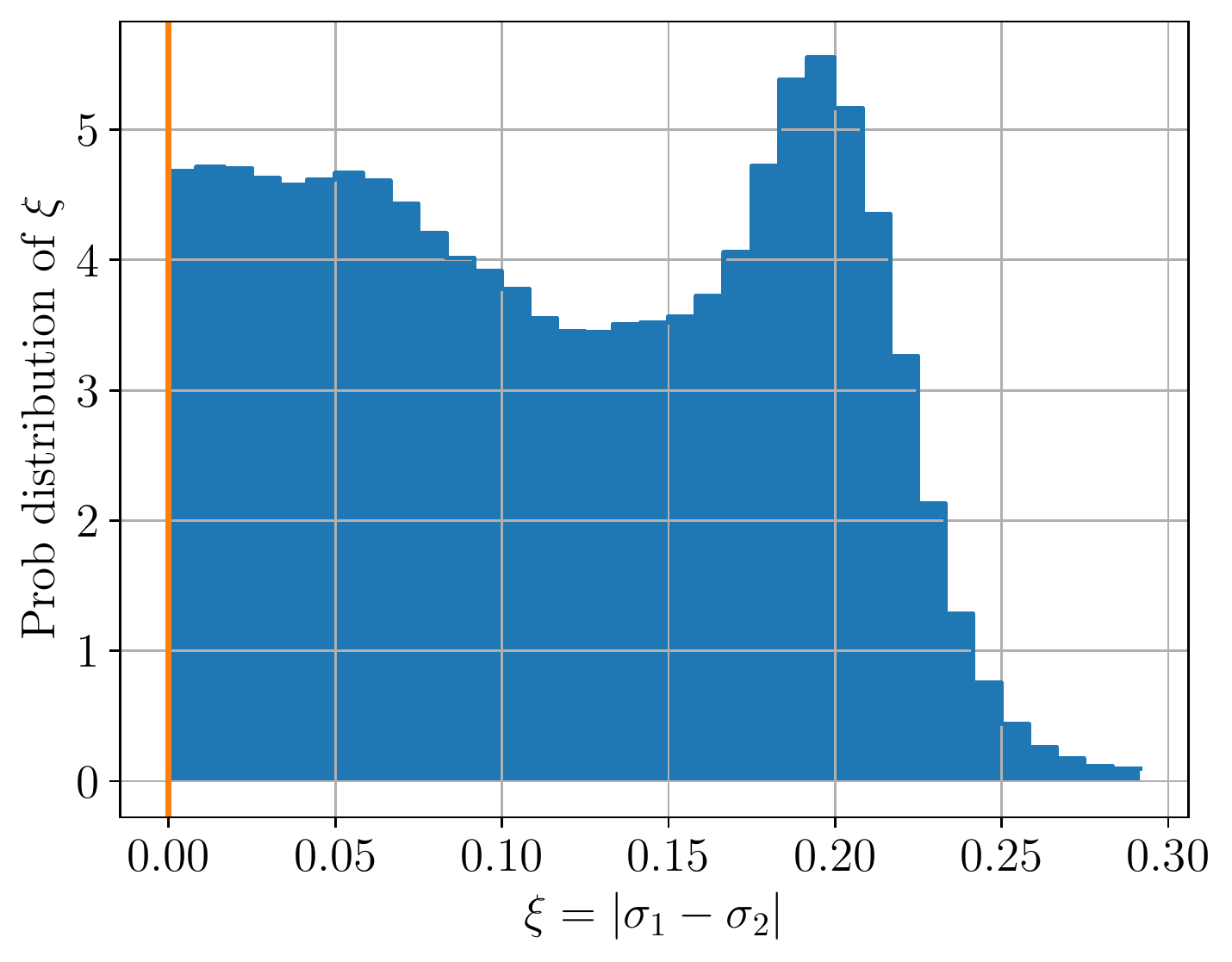}
    \caption[width = 0.8\textwidth]{Same as in Fig.~\ref{fig:00} but for 
    $(\sigma_1, \sigma_2) = (0.5, 0.5)$. 
    }  
    \label{fig:0p50p5}
\end{figure*}


We now turn to the simulations we performed to gain insight into the measurability of black hole charges
for signals typical of the long-duration binary black hole signals seen in the second 
Gravitational Wave Transient Catalog (GWTC-2) \cite{Abbott:2020niy}. Signals  
were injected into a network consisting of the two Advanced LIGO interferometers and Advanced Virgo, 
assuming stationary, Gaussian noise following the projected  
design sensitivities of these observatories   
\cite{TheLIGOScientific:2014jea,TheVirgo:2014hva}. 
As 
explained in Sec.~\ref{sec:framework}, we will focus on signals that are relatively 
low-mass, such that no more than 5\% of signal-to-noise ratio (SNR) is present beyond 
$M f = 0.018$; we require this of our injections as well. Also, we pick injected chirp masses 
$\mathcal{M} = M\nu^{3/5}$ in the range $[7, 9]\,M_\odot$, and 
mass ratios $q \in [0.4, 1]$, choices that are representative of those signals in GWTC-2 that 
satisfy our post-inspiral SNR requirement. For the purpose of studying 
the behavior of $\ln B^{\rm C}_{\rm NC}$, SNRs are chosen to be in the range 10 -- 15, again
representative of the signals in GWTC-2 that we will analyze later on. 
For $(\sigma_1, \sigma_2)$, we pick the following values:
\begin{itemize}
\item $(\sigma_1, \sigma_2) = (0, 0)$,
\item $(\sigma_1, \sigma_2) = (0.05, -0.05)$,
\item $(\sigma_1, \sigma_2) = (0.5, 0)$,
\item $(\sigma_1, \sigma_2) = (0.5, -0.5)$,
\item $(\sigma_1, \sigma_2) = (0.5, 0.5)$,
\end{itemize} 
where the larger numbers are inspired by Fisher matrix
estimates on the measurability with Advanced LIGO and Virgo of the strength of a 
dipole contribution to the phase \cite{Barausse:2016eii,Cardoso:2016olt}.

\begin{center}
\begin{table}
\begin{tabular}{ |c|c|c|c|c|c| }
\hline
$(\sigma_1, \sigma_2)$ & $(0,0)$ & $(0.05, -0.05)$ & $(0.5, 0)$ & $(0.5, -0.5)$ & $(0.5, 0.5)$ \\
\hline
$\ln B^{\rm C}_{\rm NC}$ & -4.19 & -3.11 & 10.25 & 43.82 & -1.01\\  
\hline
\end{tabular}
\caption{Values of $\ln B^{\rm C}_{\rm NC}$ for different injected values of $(\sigma_1, \sigma_2)$,
in the case of an injection with $(m_1, m_2) = (13.87, 6.36)\,M_\odot$ 
and an SNR of 12.52, for which PDFs are shown in 
Figs.~\ref{fig:00}-\ref{fig:0p50p5}.}
\label{tab:logB}
\end{table}
\end{center}

First we look at $\ln B^{\rm C}_{\rm NC}$ for 67 injections in  
stationary, Gaussian noise for an Advanced LIGO-Virgo network, 
with (dressed) masses and spins
in the ranges specified above, 
SNRs in the range 10 -- 15, and our five different choices for $(\sigma_1, \sigma_2)$. 
Histograms for the log Bayes factor are given in Fig.~\ref{fig:histograms}. The following trends 
are seen:
\begin{itemize}
\item For $(\sigma_1, \sigma_2) = (0, 0)$, all of the $\ln B^{\rm C}_{\rm NC}$ are negative 
except for one at $\ln B^{\rm C}_{\rm NC} = 0.50$, 
consistent with the absence of charges in the injected signals.
\item Also for $(\sigma_1, \sigma_2) = (0.05, -0.05)$, the great majority of $\ln B^{\rm C}_{\rm NC}$ 
are negative, indicating that 
charge-to-mass ratios of this size are not discernable at the given SNRs.
\item For $(\sigma_1, \sigma_2) = (0.5, 0)$, most signals show a positive $\ln B^{\rm C}_{\rm NC}$.
\item The choice $(\sigma_1, \sigma_2) = (0.5, -0.5)$ leads to the highest log Bayes factors, 
consistent with the fact that this yields the strongest leading-order (-1PN) contribution 
to the phasing; see Eqs.~(\ref{eq:phase}) and (\ref{eq:-1PN}). 
\item However, for $(\sigma_1, \sigma_2) = (0.5, 0.5)$, though the individual charge-to-mass ratios
are large, the -1PN contribution vanishes identically, leading to small (in fact, mostly negative) 
values of $\ln B^{\rm C}_{\rm NC}$.
\end{itemize}

Next we turn to parameter estimation. As a representative example, we focus on an injected signal
with $(m_1,m_2) = (13.87,6.36)\,M_\odot$ and an SNR of 12.52. 
Bearing in mind the invariance of our waveform model under 
$(\sigma_1, \sigma_2)\,\rightarrow\,(-\sigma_1, -\sigma_2)$, we find it convenient to 
show posteriors for $|\sigma_1|$, $|\sigma_2|$, and $\xi = |\sigma_1 - \sigma_2|$. 
 Log Bayes factors for the different injected $(\sigma_1, \sigma_2)$ 
are shown in Table \ref{tab:logB}; they are consistent with the trends observed in 
Fig.~\ref{fig:histograms}.

Fig.~\ref{fig:00} shows the results for the above mass pair and 
$(\sigma_1, \sigma_2) = (0, 0)$. We see that the posterior densities for the $|\sigma_i|$ are 
consistent with 
zero charges. They do show a peak away from zero; this is because random noise 
fluctuations cause the peaks of the distributions for the $\sigma_i$ themselves 
(before taking the absolute value) to be away from zero.
Bounds of $|\sigma_i| \lesssim 0.26$ are obtained at 68\% confidence. 
A somewhat more stringent bound is obtained for $\xi$, namely $\xi \lesssim 0.07$ 
at the same confidence level; this is again as expected because it sets the leading-order
term in the phase. 

Next, Fig.~\ref{fig:0p05n0p05} shows results for the same mass pair, but now 
$(\sigma_1, \sigma_2) = (0.05, -0.05)$. As already indicated by the log Bayes factor
in Table \ref{tab:logB}, such values of $\sigma_i$ are not detectable, and indeed the posteriors
are consistent with zero charges. However, we note that the posterior for $\xi$ does show a slight 
peak near $|\sigma_1 - \sigma_2| = 0.1$.

In Fig.~\ref{fig:0p50} we consider the case $(\sigma_1, \sigma_2) = (0.5, 0)$, for which
the log Bayes factor clearly indicated the presence of charge. Here 
the posteriors show clear support for both $(|\sigma_1|, |\sigma_2|) = (0.5, 0)$ 
and $(|\sigma_1|, |\sigma_2|) = (0, 0.5)$, consistent with another symmetry of the
waveform, namely $(\sigma_1, \sigma_2)\,\rightarrow\,(\sigma_2, \sigma_1)$. Meanwhile
the posterior for $\xi$ correctly has a strong peak near 0.5. 

Fig.~\ref{fig:0p5n0p5} shows results for $(\sigma_1, \sigma_2) = (0.5, -0.5)$. 
Though the individual posteriors for the $|\sigma_i|$ are wide, there is  
clear support for the values $(|\sigma_1|, |\sigma_2|) = (0.5, 0.5)$. The posterior for $\xi$
is tightly peaked near $\xi = 1$. 

Finally we consider the case $(\sigma_1, \sigma_2) = (0.5, 0.5)$, in Fig.~\ref{fig:0p50p5}. 
This is a case where the log Bayes factor was negative (see again Table \ref{tab:logB}), 
presumably because of the absence of the dipole contribution together with the moderate SNR. 
And indeed, the posterior for $\xi$ is not very informative, although the ones for the $|\sigma_i|$ are consistent with 
the injected values.


\section{Analysis of selected binary black hole signals}
\label{sec:realsignals}

Let us now turn to actual signals from GWTC-2 \cite{Abbott:2020niy}, and in particular those that satisfy 
our criterion that at most 5\% of the SNR resides in the post-inspiral phase, defined 
as $Mf > 0.018$. To assess which signals are in accordance with this benchmark, we 
take the median estimated parameter values reported in \cite{LIGOScientific:2018mvr,Abbott:2020niy},
and substitute them into an IMRPhenomPv2 waveform. The events we end up with are listed 
in Table \ref{tab:gwtc-2}, which also gives the values for $\ln B^{\rm C}_{\rm NC}$. Since
all log Bayes factors are negative, we find no evidence for charges in any of these.

\begin{widetext}
\begin{center}
\begin{table}
\begin{tabular}{ |c|c|c|c|c|c|c|c| }
\hline
Events & GW151226 & GW170608 & GW190707 & GW190720 & GW190728 & GW190924 & GW190930 \\
\hline
$\ln B^{\rm C}_{\rm NC}$ & -7.52 & -7.63 & -2.94 & -3.71 & -3.11 & -3.67 & -3.04 \\
\hline
\end{tabular}
\caption{The GWTC-2 events analyzed, with their log Bayes factors for charges versus no 
charges.}
\label{tab:gwtc-2}
\end{table}
\end{center}
\end{widetext}

\begin{figure*}
    \centering
   \includegraphics[keepaspectratio, width=6cm]{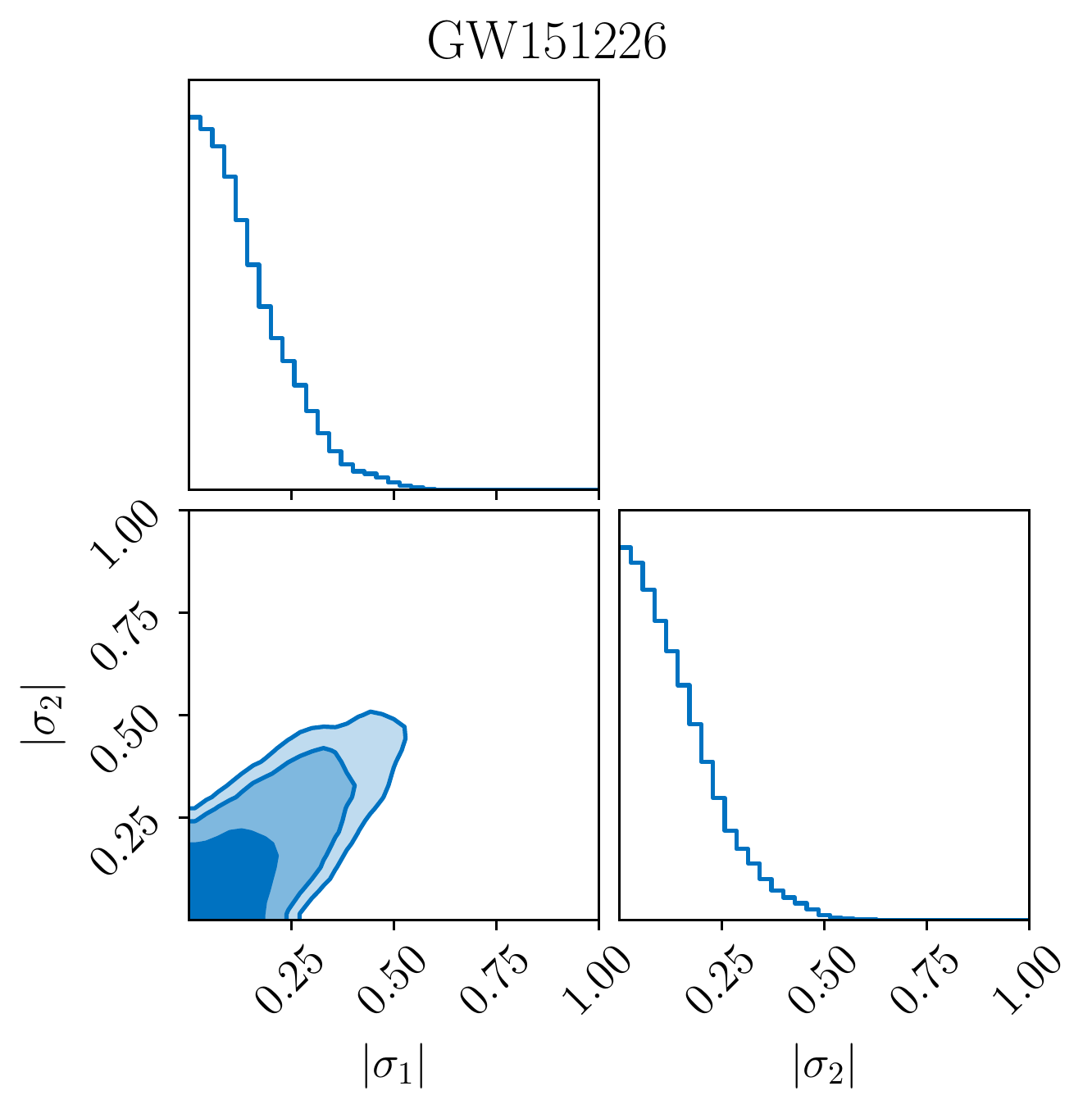}
   \includegraphics[keepaspectratio, width=6cm]{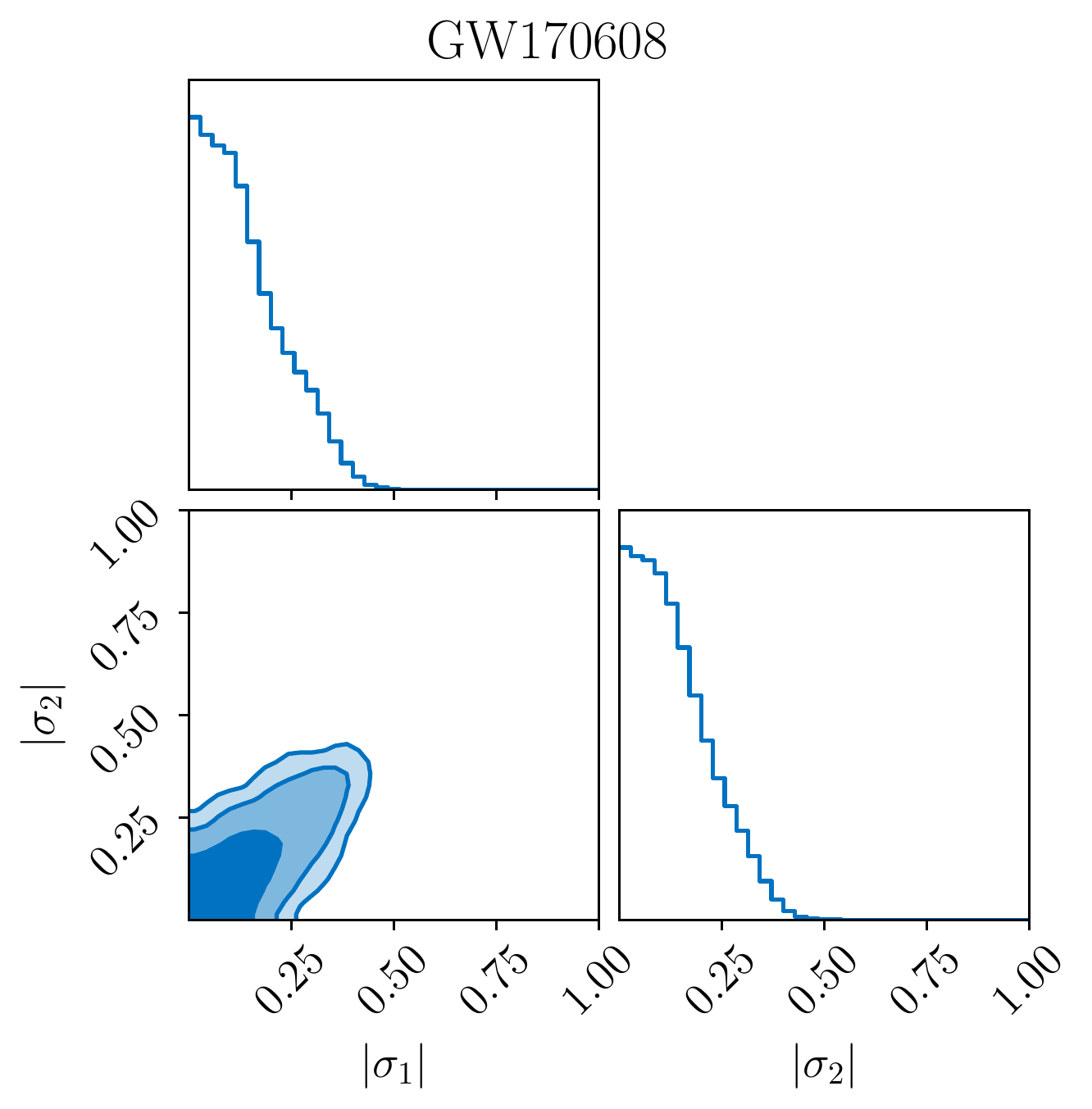}
   \includegraphics[keepaspectratio, width=6cm]{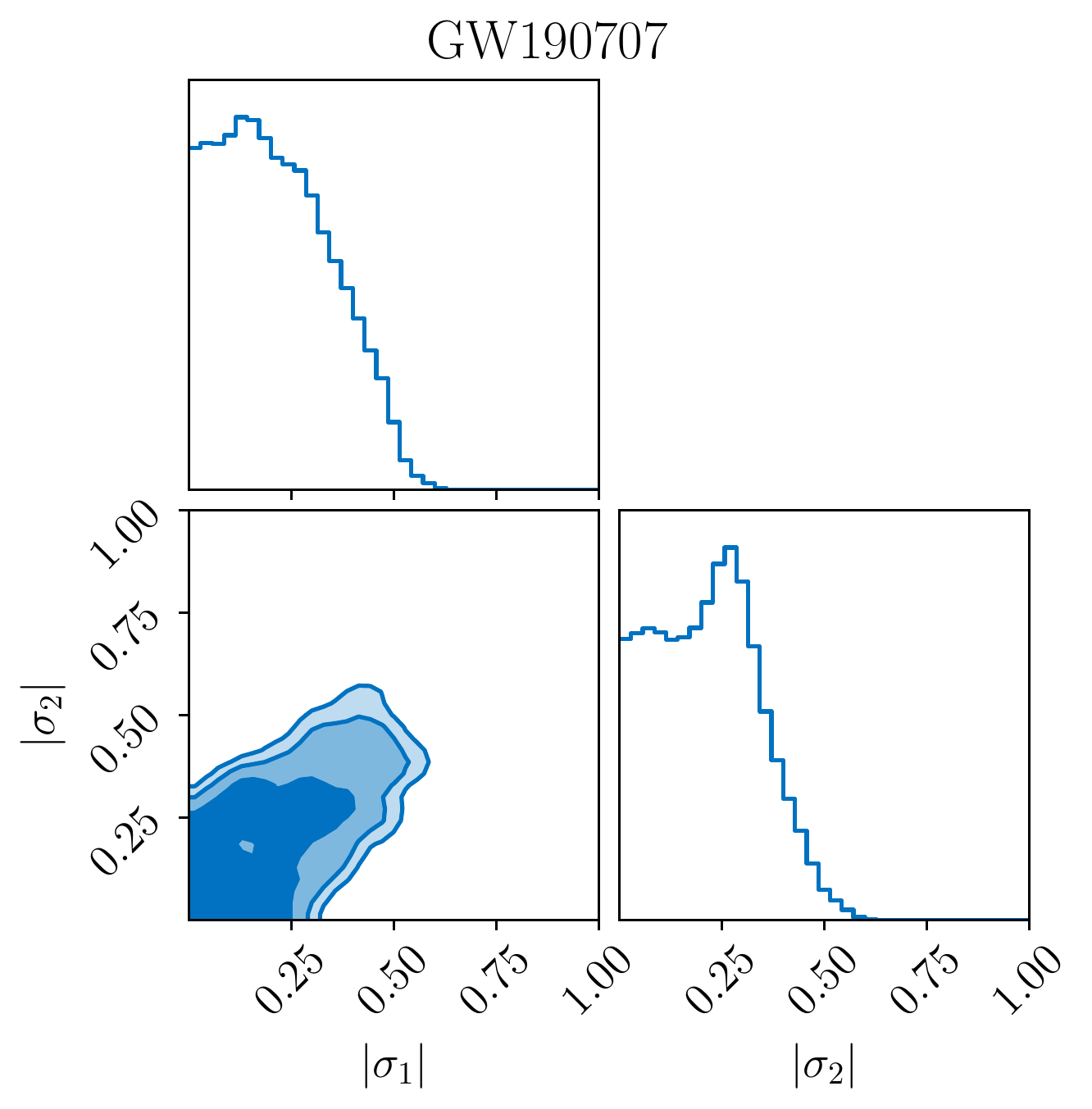}
   \includegraphics[keepaspectratio, width=6cm]{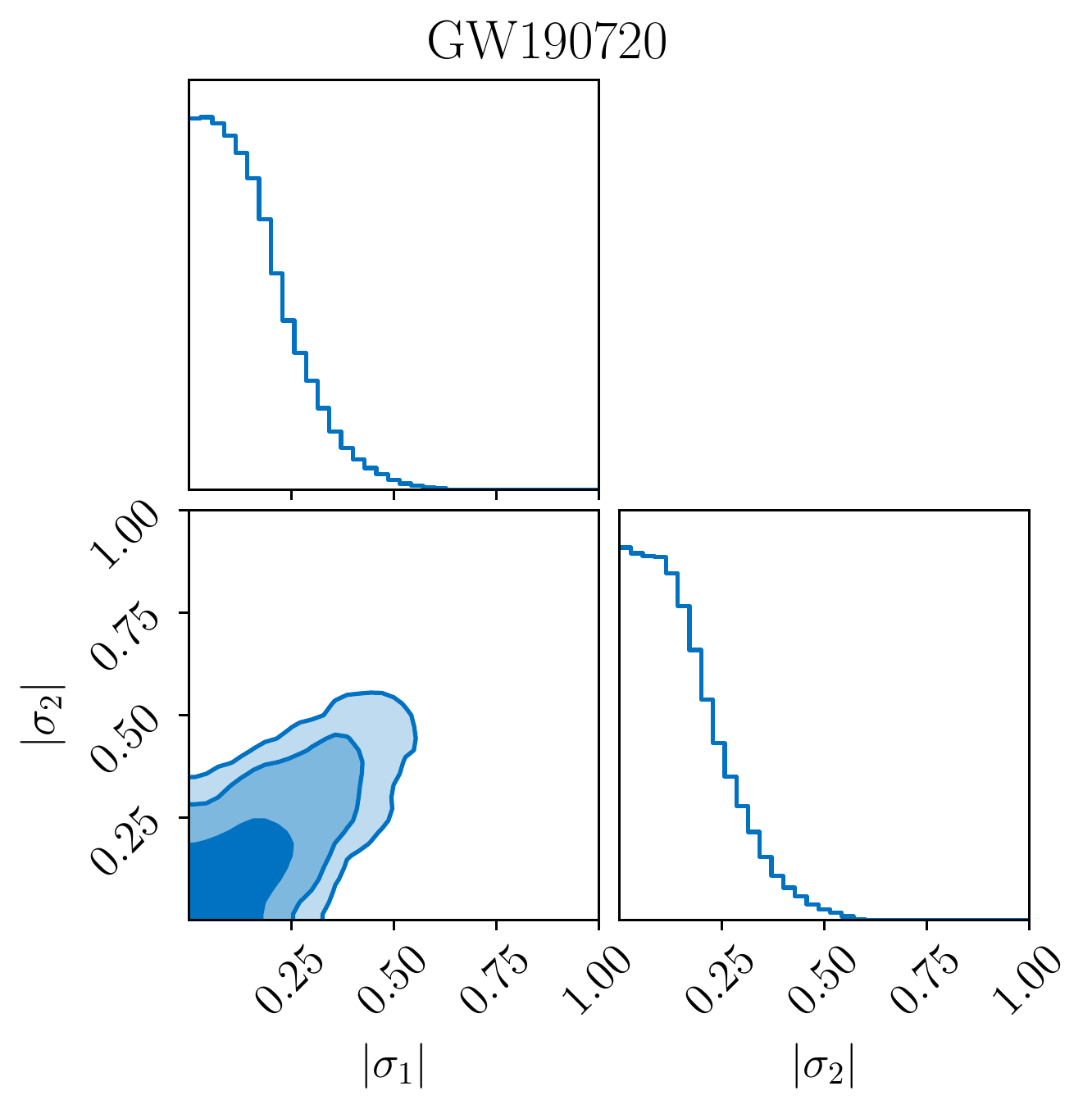}
   \includegraphics[keepaspectratio, width=6cm]{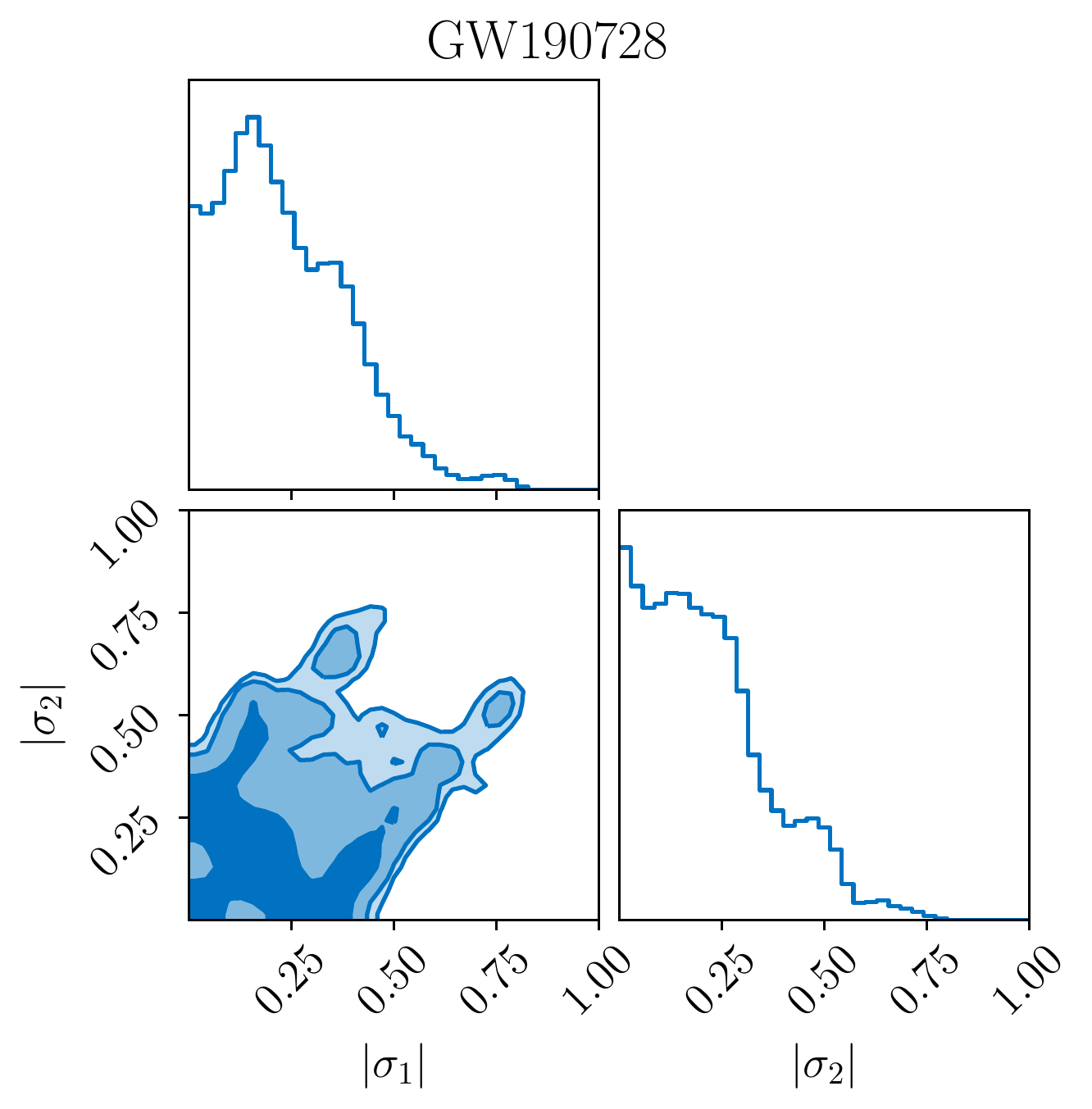}
   \includegraphics[keepaspectratio, width=6cm]{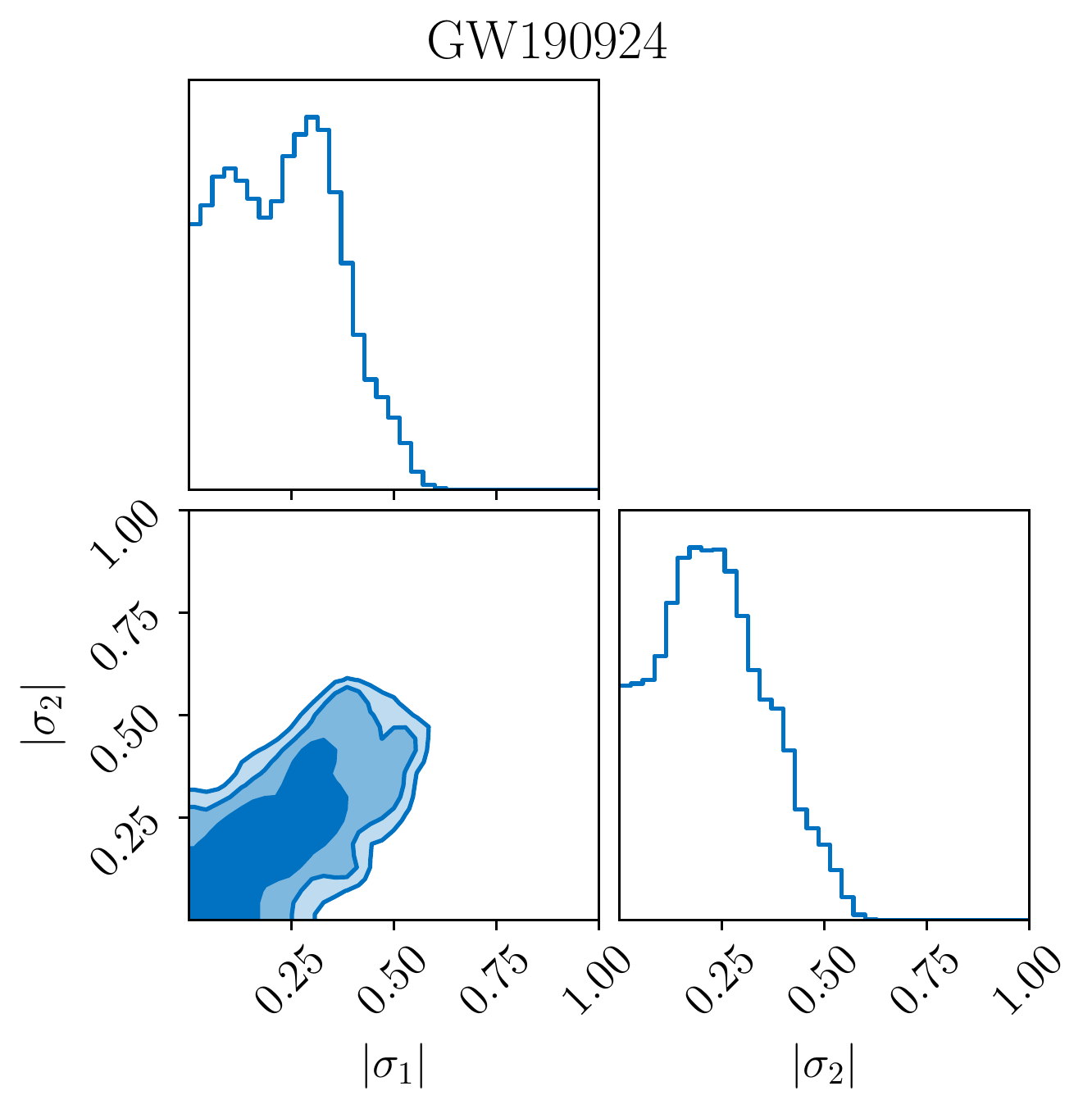}
   \includegraphics[keepaspectratio, width=6cm]{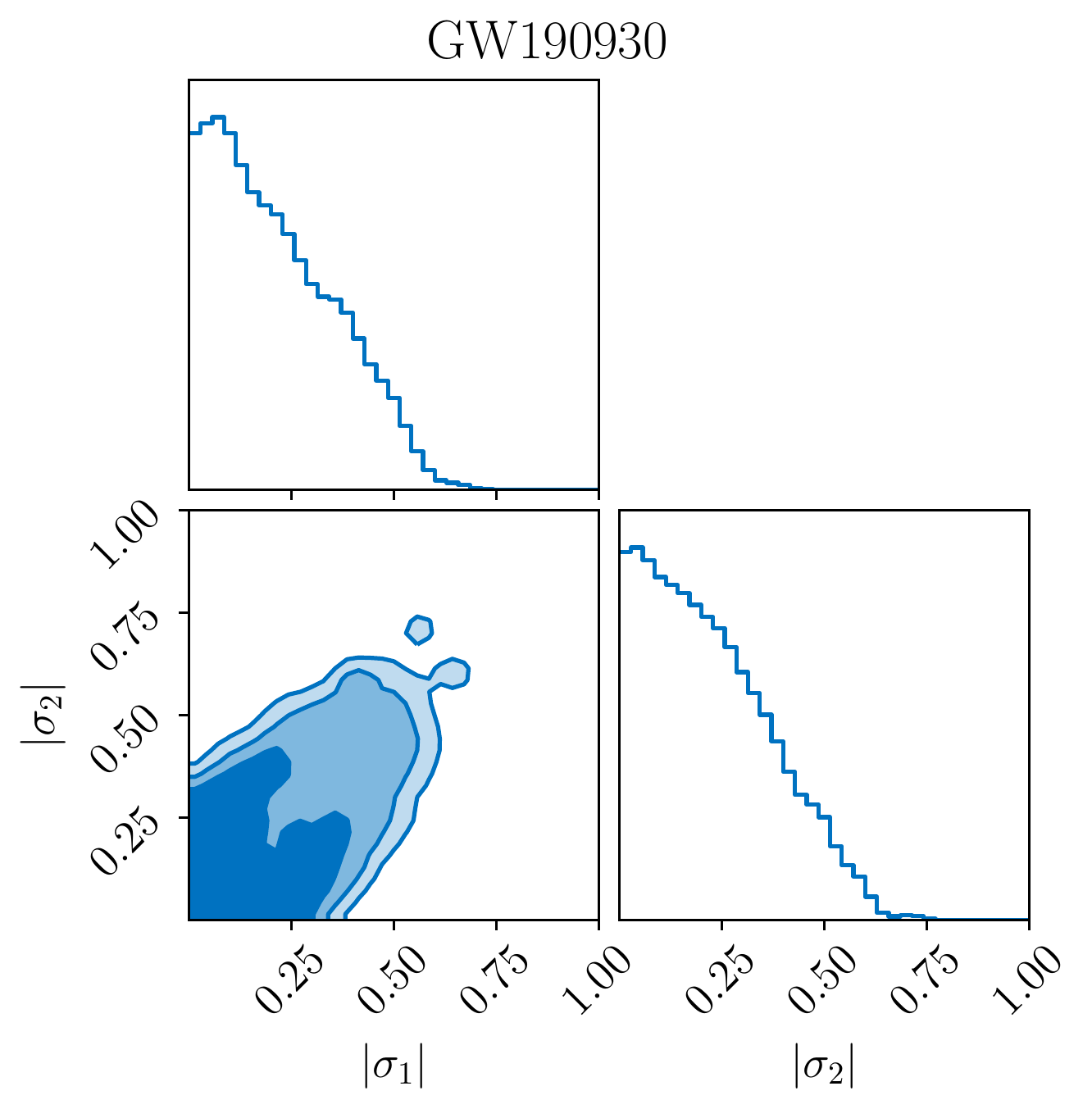} 
    \caption[width = 0.8\textwidth]{
    Corner plots for the posteriors of $|\sigma_1|$, $|\sigma_2|$, for the events of Table 
    \ref{tab:gwtc-2}.
    }  
    \label{fig:sigmas}
\end{figure*}

\begin{figure*}
    \centering
   \includegraphics[keepaspectratio, width=15.0cm]{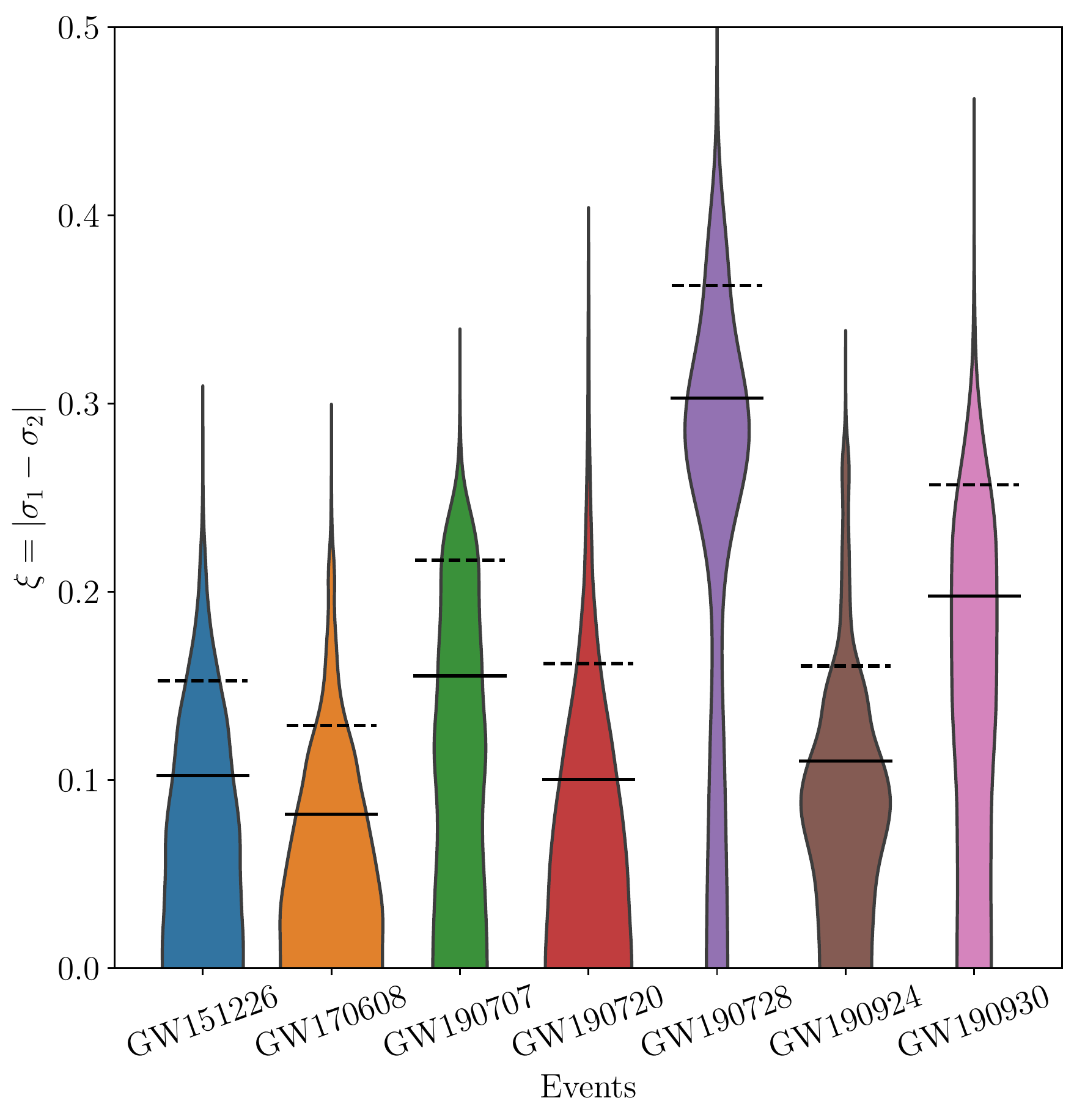}
    \caption[width = 0.8\textwidth]{Posterior densities for $\xi = |\sigma_1 - \sigma_2|$, 
    for the events of Table \ref{tab:gwtc-2}. The solid lines indicate 68\% confidence levels, 
    the dashed lines 90\% confidence levels. 
    }  
    \label{fig:xi}
\end{figure*}

For completeness, we also show posteriors for $|\sigma_1|$ and $|\sigma_2|$ (Fig.~\ref{fig:sigmas}), 
and $\xi = |\sigma_1 - \sigma_2|$ (Fig.~\ref{fig:xi}). Here too, all the signals 
show consistency with $(\sigma_1, \sigma_2) = (0,0)$. Events like GW190707, 
GW190728, and GW190924 have posteriors for the individual $|\sigma_1|$ and $|\sigma_2|$ that seem 
to have a peak away from zero, but as in the case of our simulated signal with 
$(\sigma_1, \sigma_2) = (0,0)$ (see Fig.~\ref{fig:00}), this can be attributed 
to noise fluctuations causing the peaks of the $\sigma_i$ themselves (before taking the 
absolute value) to be away from zero.
However, these three events also have a peak in  $\xi$ that is away from zero; in the case 
of GW190728 there is even a relatively strong peak at $\xi \sim 0.3$. 
That said, the log Bayes factor for GW190728 ($\ln B^{\rm C}_{\rm NC} = -3.11$) is below
the largest log Bayes factor for injections with $(\sigma_1, \sigma_2) = (0,0)$ 
shown in Fig.~\ref{fig:histograms}, which is $\ln B^{\rm C}_{\rm NC} = 0.50$; the same 
is true of all the other real events in Table \ref{tab:gwtc-2}. Although the  
injection set of Fig.~\ref{fig:histograms} pertained to stationary, Gaussian noise, we expect a more complete 
``background distribution'' for $\ln B^{\rm C}_{\rm NC}$ in real noise to extend to even 
larger values. Therefore, we are not induced to conclude that charges were present on any of the 
binary black holes that generated the real signals we analyzed.

For all our real events, the 1-$\sigma$ bounds on 
the $|\sigma_i|$ tend to be at the level of $0.2$ -- $0.3$, consistent with the zero-charge injection 
which we studied PDFs for in the previous section. Similarly, bounds on $\xi$ tend to be somewhat 
more stringent, varying from $0.08$ (for GW170608) to $0.3$ (for GW190728).

\section{Summary and conclusions}
\label{sec:conclusions}

We have set up a Bayesian analysis framework to search for, or constrain, (dark) electric
charges on binary black holes using gravitational waves. In particular, the inspiral 
part of the phasing of the precessing-spin IMRPhenomPv2 inspiral-merger-ringdown waveform was modified 
to include the effect of such charges up to 1PN order. This was then used for both injections 
and template waveforms, focusing on signals with less than 5\% of their SNR in the post-inspiral
regime, in view of the currently unknown effect of charges during plunge and merger. 

To test the analysis set-up, we looked at the log Bayes 
factor, $\ln B^{\rm C}_{\rm NC}$, comparing the hypothesis that charges are present with 
the one that assumes zero charges, for signals with SNRs between 10 and 15. Choosing different injected values for the charge-to-mass 
ratios $(\sigma_1, \sigma_2)$, expected trends were seen in the distributions of $\ln B^{\rm C}_{\rm NC}$:
(a) when the $\sigma_i$ were zero or small, the great majority of our simulated signals yielded 
$\ln B^{\rm C}_{\rm NC} < 0$, and (b) for larger $\sigma_i$, the typical magnitude of 
$\ln B^{\rm C}_{\rm NC}$ was set by the strength of the leading-order contribution of
charges to the phase, which is determined by $\xi = |\sigma_1 - \sigma_2|$. 

As a case study for parameter estimation we used an injection with an SNR of 12.52. 
PDFs were indicative of the injected $(\sigma_1, \sigma_2)$,
and for $(\sigma_1, \sigma_2) = (0,0)$, 1-$\sigma$ upper bounds came out to be 
$|\sigma_i| \lesssim 0.26$ and $\xi \lesssim 0.07$.

Finally, we turned to real signals from GWTC-2, again selected to have a long inspiral in band.
All of the $\ln B^{\rm C}_{\rm NC}$ came out to be negative, consistent with the absence of
charges, and also the PDFs for the $|\sigma_i|$ and $\xi$ were consistent with zero charge. 
Typical bounds on charge-related parameters were $|\sigma_i| \lesssim 0.2 - 0.3$ and 
$\xi \lesssim 0.08 - 0.3$. 

In this work we focused on the inspiral regime, but charge-induced modifications of
the ringdown spectrum have also been computed 
\cite{Pani:2013ija,Pani:2013wsa,Zilhao:2014wqa,Mark:2014aja,Dias:2015wqa}. It would be
of interest to search for the signature of charges in the ringdown signal of high-mass events,
whose ringdown modes are starting to be probed even with Advanced LIGO and Virgo at 
O3 sensitivity \cite{LIGOScientific:2020tif}. Finally, should appropriate waveform models become 
available in the future, it will be interesting to see how charge measurements will sharpen when 
the entire inspiral-merger-ringdown process can be used.

\begin{acknowledgments} 

  P.K.G., P.T.H.P., G.K., and C.V.D.B.~are supported by the research programme 
  of the Netherlands Organisation for Scientific Research (NWO). 
  The authors are grateful for computational resources provided by the 
  LIGO Laboratory and supported by the National Science Foundation Grants No.~PHY-0757058 and 
  No.~PHY-0823459. 
  This
  research has made use of data, software and/or web tools
  obtained from the Gravitational Wave Open Science Center (https://www.gw-openscience.org), a service of LIGO
  Laboratory, the LIGO Scientific Collaboration and the
  Virgo Collaboration. LIGO is funded by the U.S. National Science Foundation. Virgo is funded by the French
  Centre National de Recherche Scientifique (CNRS), the
  Italian Istituto Nazionale della Fisica Nucleare (INFN)
  and the Dutch Nikhef, with contributions by Polish and
  Hungarian institutes.
  
\end{acknowledgments}

\bibliography{refs}

\onecolumngrid

\end{document}